\pdfoutput=1
\documentclass[useAMS,usenatbib,letterpaper]{mn2e}
\usepackage{graphicx}
\usepackage{rotating}
\usepackage{lscape}

\title[The Cepheids of NGC1866]{The Cepheids of NGC1866: A Precise Benchmark for the Extragalactic Distance Scale and Stellar Evolution from Modern {\it UBVI\/} Photometry}
\author[Musella et al.]{I. Musella$^{1}$\thanks{E-mail:
ilaria.musella@oacn.inaf.it}, M. Marconi$^{1}$, P. B. Stetson$^{2}$,
G. Raimondo$^{3}$, E. Brocato$^{4}$, R. Molinaro$^{1}$, 
\newauthor
V. Ripepi$^{1}$, R. Carini$^{4}$, G. Coppola$^{1}$, A. R. Walker$^{5}$, D. L. Welch$^{6}$\\
$^{1}$ INAF-Osservatorio Astronomico di Capodimonte, Via Moiariello 16, I-80131 Napoli, Italy\\
$^{2}$ Dominion Astrophysical Observatory, NRC-Herzberg, 5071 West Saanich Road, Victoria, BC V9E 2E7, Canada\\
$^{3}$ INAF-Osservatorio Astronomico Collurania, via M. Maggini,
I-64100 Teramo, Italy\\
$^{4}$ INAF-Osservatorio Astronomico di Roma, Via Frascati 33, I-00044 Monte Porzio Catone, Italy\\
$^{5}$ Cerro Tololo Inter-American Observatory, National Optical Astronomy Observatory, Casilla 603, La Serena, Chile\\
$^{6}$ Department of Physics \& Astronomy, McMaster University, Hamilton, Ontario, L8 S 4M1, Canada
}

\begin{document}

\date{Accepted ..... Received .... ;}

\pagerange{\pageref{firstpage}--\pageref{lastpage}} \pubyear{0000}

\maketitle

\label{firstpage}

\begin{abstract}
We present the analysis of multiband time-series data for a
sample of 24 Cepheids in the field of the Large Magellanic Cloud
cluster NGC1866. Very accurate {\it BVI\/} VLT photometry is combined with
archival {\it UBVI\/} data, covering a large temporal window, to obtain
precise  mean magnitudes and periods with typical errors of 1-2\% and
of 1 ppm, respectively. 
These results represent the first accurate and homogeneous dataset 
 for a substantial sample
of Cepheid variables belonging to a cluster and hence sharing common
distance, age and
original chemical composition. 
Comparisons of the resulting multiband Period-Luminosity and
Wesenheit relations to both empirical and theoretical results for
the Large Magellanic Cloud
are presented and discussed to derive the distance
of the cluster and to constrain the mass-luminosity relation of the
Cepheids.
The adopted theoretical scenario is also tested by comparison with
independent calibrations of the Cepheid Wesenheit zero point based on
trigonometric parallaxes and Baade-Wesselink techniques.
Our analysis suggests that a mild
overshooting and/or a moderate mass loss can affect intermediate-mass stellar evolution in
this cluster and gives a distance modulus of $18.50 \pm 0.01$ mag.
The obtained {\it V,I\/} color-magnitude diagram is also analysed and compared
with both synthetic models and theoretical isochrones for a range of
ages and metallicities and for different efficiencies of core overshooting. 
As a result, we find that the age of NGC1866 is about 140 Myr,
assuming $Z=0.008$ and the mild efficiency of overshooting suggested by
the comparison with the pulsation models. 

\end{abstract}


\section{Introduction} \label{introduction}

Classical Cepheids play a fundamental role in the calibration of the extragalactic
distance scale thanks to their characteristic period-luminosity (PL)
and period-luminosity-color (PLC) relations. In particular, the
application of  a Large Magellanic Cloud (LMC)-based PL relation to
external galaxies observed with the Hubble Space Telescope (HST) has
led to the calibration of secondary distance indicators and in turn to
an estimate of the Hubble constant \citep[H$_0$; see,][and
references therein]{Freedman01,Saha01,Sandage06,Riess11}.

The LMC distance has traditionally played a crucial role in the extragalactic distance
determination, with values lower than 18.50 mag implying the so called
``short distance scale'' and values larger than 18.50 mag suggesting a
``long distance scale''. 

One of the most important issues to be considered when dealing with the
Cepheid PL relation is its dependence on the metallicity.  The
universality of the PL relation, and the possibility that its slope and/or
zero-point might depend on the chemical composition, have been actively
debated for two decades, on both observational and theoretical grounds,
\citep[see
e.g.][]{Gould94,Sasselov97,Kennicutt98,Fiorentino02,Sakai04,Marconi05,Bono08,Romaniello08,Marconi10,Sandage06,FM11,Gerke11,Fausnaugh15},
with controversial results. No general consensus has been reached so
far.  Empirical calibrations of the LMC PL and PLC relations are
generally based on field Cepheids, implying the presence of uncertainties due to the
range in distance \citep[depth effects][]{vdm01},
metallicity \citep[see e. g.][and references therein]{Davies15,Luck98}
and
extinction \citep[differential reddening, see e.g.][and references therein]{Haschke11}.

On the other hand, any theoretical scenario for pulsation models aimed at
providing robust support to empirical calibrations needs to be
tested against observational constraints based on statistically significant samples
of Cepheids with accurate light curves.

The populous blue LMC cluster NGC1866 is already
known to host an exceptionally rich sample of  
more then 20 Cepheids \citep[][]{WS93,Musella06}. One of these
was identified by \citet{Musella06} in a preliminary
analysis of the proprietary {\it BVI\/} VLT data.
It is unquestionable that such a unique sample of
Cepheids---likely all members of the cluster and at
the same distance, chemical composition and age---would 
constitute a milestone in our understanding of the Cepheid
pulsational scenario. Indeed, it offers an unprecedented opportunity to
investigate both empirical and theoretical estimates of the
luminosity and color of the pulsating structures and their
relation with the observed periods. 
For this reason, many authors have studied the NGC1866 Cepheids
\citep{Welch91,WS93,Gieren94,Walker95,Gieren00,Storm05,Testa07} 
in both the optical and near infrared bands, and have tested different
methods to calibrate the PL relations in different filters.  
In \citet{Brocato04} we already discussed the sample of the 23 known
Cepheids in NGC1866, concluding that unfortunately only 4 to 6 Cepheids  
had light curves accurate enough to allow a meaningful
determination of their luminosities and colors. 
On the basis of such a tantalizing situation, we took advantage of
assigned observing time at the ESO Very Large Telescope to perform an accurate photometric investigation of the cluster field, with the aim of
securing suitable data constraining the light curves of all the member
Cepheids. Moreover, to get accurate information about
radial velocities and 
chemical abundances of the stars in NGC1866, we have performed FLAMES@VLT
spectroscopic observations for 30 stars (19 belonging to the cluster
and 11 to the LMC field), including 3 Cepheids
\citep{Mucciarelli11,Molinaro12}. \citet{Mucciarelli11} found that, as
far as the chemical composition is concerned, the cluster stars are reasonably homogeneous.
Indeed, they appear to share the same abundances within the uncertainties, and
this property is 
independent of the evolutionary status. The average iron
abundance is $[Fe/H]=-0.43 \pm 0.01$ dex, with a dispersion $\sigma =
0.04$ dex.  For the three spectroscopically investigated Cepheids \citet{Molinaro12}, adopting the same procedure used in
\citet{Mucciarelli11}, found values fully consistent with the average
iron content. 
Moreover, \citet{Molinaro12} applied the CORS Baade-Wesselink method \citep{Ripepi97}
to a sample of 11 Cepheids,
using radial velocities obtained both from our FLAMES investigation
and from literature data (see references therein), and light curves
based on a part of the
{\it UBVI\/} data used in this paper complemented with {\it K\/} data by
\citet{Testa07}.  In this way, they obtained  a direct estimate of the distance
modulus of NGC1866,  $\mu_0 =18.51 \pm 0.03$ mag
\citep[see][for details]{Molinaro12}. 

In this paper, we analyze the properties of NGC1866 Cepheids, 
relying on an extensive multifilter dataset not only to derive
information about the distance, but also to test pulsational and
evolutionary theoretical models.  

 The  paper is organized as follows. We present observations and data reduction in
 section \ref{data}  and Cepheid properties in section \ref{cepheids}.
The adopted pulsational models are briefly described in
section \ref{models}. To derive the distance of NGC1866 we have
applied in section \ref{relations} the theoretical
period-luminosity and  Wesenheit relations, and in section \ref{distance}
two empirical calibrations based on
trigonometric parallaxes and the Baade-Wesselink method. 
In section \ref{cmd}, a deep and accurate {\it V,V-I\/} color-magnitude diagram is
presented and 
compared with theoretical isochrones to derive additional information
on the cluster properties. The conclusions close the paper.

\begin{figure}
\includegraphics[width=8.5cm]{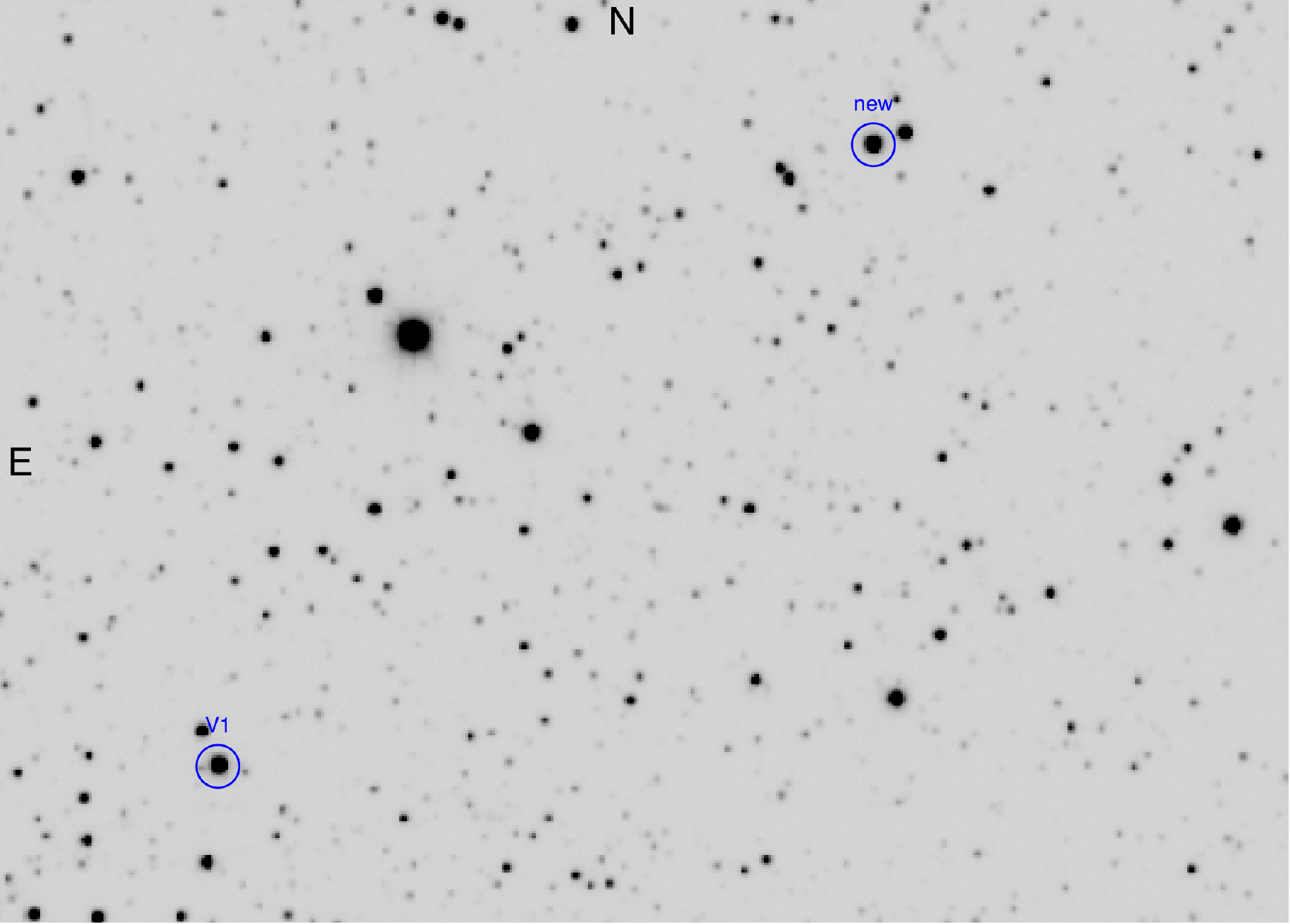} 
\caption{Identification map (a region of $2 \times 1.5$ arcmin) for the newly discovered
Cepheid in the cluster NGC1866.} \label{fig:map}
\end{figure}

\section{Observations and data reduction.} \label{data}

Our photometric database comprises 1471 CCD images (see  Table \ref{log}). 

\begin{table*}
\caption{Log of the Observations, Instrumental Setup and number of
   images obtained in each filter.} \label{log}
\begin{center}
\begin{tabular}{|c|c|c|c|c|c|c|}
\hline
\hline
\noalign{\smallskip}
Dates                  & Telescope           & Detector & $N_U$   &
$N_B$   & $N_V$   & $N_I$         \\
\noalign{\smallskip}
\hline
\noalign{\smallskip}
1986 Dec 18-23              & CTIO 0.9m           & RCA          & 16  & 42  & 37  & 21  \\
1987 Nov 22-23              & CTIO 4.0m           & RCA1         & 3   & 17  & 18  & --- \\
1987 Nov 30-Dec 01          & CTIO 0.9m           & TI1          & 1   & 1   & 1   & --- \\
1988 Jan 17-19              & CTIO 4.0m           & TI1          & 2   & 2   & --- & --- \\
1990 Sep 11                 & CTIO 4.0m           & TI1          & 5   & 5   & 5   & 5   \\
1990 Oct 25-26              & CTIO 0.9m           & TI3          & 6   & 6   & 6   & 6   \\ 
1990 Dec 13                 & CTIO 4.0m           & Tek512       & --- & 5   & 6   & 6   \\
1990 Dec 16-17              & CTIO 0.9m           & Thomson 1024 & 4   & 6   & 9   & 4   \\
1991 Jan 22                 & CTIO 1.5m           & Tek1024      & 2   & 3   & 3   & 3   \\
1991 Feb 01                 & CTIO 0.9m           & Tek1024      & --- & 2   & 2   & 2   \\
1991 Mar 03                 & CTIO 4.0m           & Thomson 1024 & --- & 6   & 6   & 6   \\
1993 Jan 14-2000 Jan 03$^*$ & Mt. Stromlo 50-inch & CCD3 \& CCD4 & --- & --- & 309 & --- \\
1993 Nov 02-05              & CTIO 0.9m           & Tek1024      & --- & 2   & 2   & --- \\
1994 Nov 24-28              & CTIO 0.9m           & Tek2K\_4     & --- & 13  & 13  & 13  \\
1999 Dec 01-04              & CTIO 4.0m           & Mosaic2      & --- & 44  & 44  & 46  \\
2001 Jan 17-19              & CTIO 4.0m           & Mosaic2      & --- & --- & 8   & 8   \\
2001 Mar 23-30              & MPI/ESO 2.2m        & WFI          & --- & 24  & 16  & 24  \\
2001 Apr 06-12              & Mt. Stromlo 74-inch & CCD17        & --- & 29  & 38  & 34  \\
2001 Dec 12-13              & CTIO 1.5m           & Site2K\_6    & --- & 1   & 1   & 1   \\
2003 Oct 03-Dec 25          & ESO VLT 8.0m        & FORS1        & --- & 69  & 90  & 62  \\
2008 Aug 26-28              & CTIO 4.0m           & Mosaic2      & --- & --- & 16  & 16  \\
\hline
\multicolumn{7}{l}{\footnotesize{$^*$ MACHO data}}
\end{tabular}
\end{center}
\end{table*}

The median seeing of the images was 1.8 arcsec, 
but it is worth noting that for the {\it B\/}, {\it V\/}, and {\it I\/} filters, we also
have a set of high precision
photometric proprietary data obtained with FORS1@VLT. 
For this data set, the detector was a 2048x2048 Tektronix
CCD with $24\mu \times 24\mu$ pixels. Projected on the sky,
the scale is 0.2 arcsec/pixel for a total field of view of
$6.8^{'}\times6.8^{'}$. We have obtained time-series photometry on one pointing centered on
NGC1866 in the {\it BVI\/} filters with exposures of 60 s in
each filter. The observations were efficiently carried out in
service mode, with the constraint of seeing better than
$0.7^{''}$ because the target cluster is very crowded and many
Cepheids are located near the cluster center. 

The reduction and calibration of  all the photometric data was carried out with the DAOPHOT/ALLFRAME
packages \citep[][]{Stetson87,Stetson94}---which combine excellent precision
with a large degree of automation---and 
using standard
methodologies as described in, for example, \citet{Stetson00} and \citet{Stetson05}. Special care has been taken in deriving an
accurate PSF for each image because of the high degree of crowding
not only in the central regions of the cluster but also in its
outskirts. 

The calibration was based on local standards in NGC1866. The
median internal photometric precision based on these local standards is
0.01, 0.001, 0.0008 and 0.002 mag for {\it U\/}, {\it B\/}, {\it V\/} and {\it I\/} bands,
respectively, using all the data sets, and 0.0006, 0.0004 and
0.0008 mag for the {\it B\/}, {\it V\/} and {\it I\/} bands, respectively, 
considering only the VLT data. Concerning the external accuracy of our 
photometry, the root-mean-square differences of our mean magnitudes for
Landolt's standards \citep{Landolt92}, on a star-by-star basis, is of $\sim
0.029$, $\sim
0.016$, $\sim 0.013$ and $\sim 0.016$ mag in the {\it U\/}, {\it B\/}, {\it V\/} and {\it I\/} bands, respectively. These differences represent 180, 378, 400, and 250 individual Landolt standards, so our {\it mean\/} photometric system differs from
Landolt's {\it mean\/} system by not less than $\sim 0.001$~mag in {\it B\/},{\it V\/} and {\it I\/}, and $\sim 0.002$~mag in {\it U\/}. This is
probably a fundamental limit for the external, absolute accuracy that
can be obtained with CCD measurements, given that we use different filters and
detectors than Landolt used.

\section{NGC1866 Cepheids } \label{cepheids}

\begin{figure}
\includegraphics[width=8.5cm]{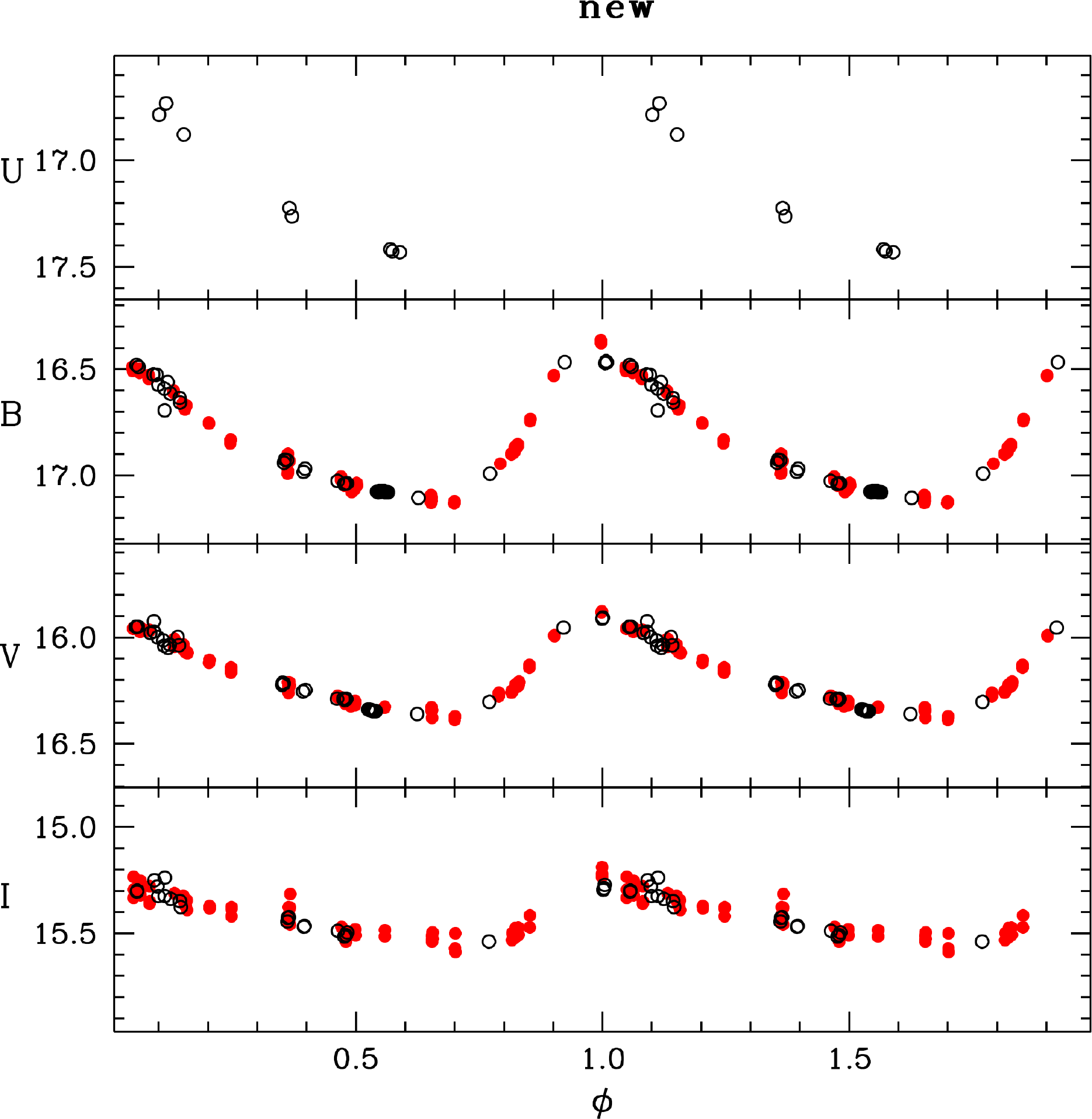} 
\caption{Light
curves of the newly discovered Cepheid in NGC1866. Marks are the same
as used in Fig. \ref{fig:lcurves}.} \label{fig:newcef} 
\end{figure}

The position and the light curves in the {\it UBVI\/} bands of the Cepheid identified in \citet{Musella06} are
reported in Figs. \ref{fig:map} and \ref{fig:newcef},
respectively. The coordinates of this Cepheid are $\alpha=$
05:13:20.79 and $\delta=-$65:24:57.8 (J2000) at about 3.5 arcmin from
the cluster center.
The light curves of all other previously known variables are plotted in Fig. \ref{fig:lcurves} In particular,
grey dots are data from the MACHO project, red dots are VLT data and open circles are all the other
archival data (see Table \ref{log}). 

The location in the extreme cluster
periphery of the Cepheid  identified in \citet{Musella06} 
may raise serious doubts about its membership. 
\citet{Lupton89} suggest that NGC1866 is not
  tidally limited but is embedded in an unbound halo. \citet{Noyola07},
  imposing the existence of a tidal radius, find an half-light radius
  of 49.7 arcsec: much less than the distance of this Cepheid ($\sim
  3.4$ arcmin).
However, in the
following we will demonstrate that both the
location in the  color-magnitude diagram (CMD) and the pulsational
properties of this variable appear in close agreement with the
behaviour of the other
cluster Cepheids. Obviously, to have a firmer conclusion about the
membership of this distant Cepheid, the mean
radial velocity should also be measured through accurate spectroscopic observations.

The obtained light curves show the superior accuracy of the VLT
photometry when compared with the other datasets. In particular, for the MACHO
data in the {\it V\/} band, it was necessary to reject several
scattered phase points. We use all datasets, spanning
a large time window, to determine accurate periods using
Period04 \citep{Lenz05}. This program allows us to determine also the
error on the derived period via 
Monte Carlo simulations. On the other hand,
to obtain mean (intensity-averaged) magnitudes and colors as accurately
as possible, we fitted the light curves in all available bands with a
smoothing spline obtained with a C code written by one of the authors
(R. M).  In particular,
for the {\it B\/}, {\it V\/} and {\it I\/} light curves we used only the VLT data, adopting the other
datasets only for the {\it U\/} filter. 
All the derived periods, {\it UBVI\/} mean
magnitudes and amplitudes $A_{\lambda}$ (for each $\lambda$ filter) are reported in Table \ref{magcef}, but we will not use the
{\it U\/} band in the following analysis. 
As an estimate of the uncertainty in the calculated mean magnitudes, we report in
Table \ref{magcef} the r.m.s. of the residuals of the data around the fitted curves; of course, these residuals will not fully represent consistent photometric errors resulting from the blending conditions of individual stars. 
Table \ref{magcef} also reports the complementary {\it K\/} band data
obtained by \citet{Testa07}.

\begin{table*}
\caption{Properties of the Cepheids in NGC1866.} \label{magcef}
\begin{center}
\begin{tabular}{|c|c|c|c|c|c|c|c|c|c|c|}
\hline
\hline
\noalign{\smallskip}
ID & {\it P\/}  & {\it U\/} (r.m.s.) & {\it B\/}  (r.m.s.) & {\it V\/} (r.m.s.) & {\it I\/} (r.m.s.) & {\it K\/}  & $A_U$ & $A_B$ & $A_V$ & $A_I$\\
 & [days]  & [mag] & [mag] & [mag] &[mag] & [mag]  & [mag] & [mag] & [mag] & [mag]\\
\noalign{\smallskip}
\hline
\noalign{\smallskip}

V6       & $  1.944262  \pm  0.000001  $ & 16.90 (0.04) & 16.76 (0.01)
& 16.18 (0.01) & 15.49 (0.01) & 14.60 &  0.40 & 0.34 & 0.25 & 0.15\\
V8       & $  2.007     \pm     0.003  $ & 16.94 (0.04) & 16.80 (0.01)
& 16.19 (0.01)& 15.47 (0.02)& 14.57 & 0.60 & 0.46 & 0.26 & 0.20\\
V5$^{**}$       & $  2.039071     \pm     0.000004  $ & --- & 15.93 (0.02) &15.60 (0.02)& 15.14 (0.02)& --- & --- & 0.32 & 0.26 & 0.18\\
HV12199  & $  2.63916   \pm    0.00001 $ & 17.10 (0.10) & 16.90 (0.03) &
16.29 (0.02)& 15.57 (0.02)& 14.70  & 1.13 &0.99 & 0.70 & 0.41\\
HV12200  & $  2.72498   \pm    0.00002 $ & 16.92 (0.06)& 16.74 (0.02) &
16.16 (0.02) & 15.48 (0.02) & --- & 1.49 & 1.18 & 0.82 & 0.51 \\
We4$^*$       & $  2.86036   \pm    0.00002 $ & 16.52 (0.09) & 16.44 (0.04) &
15.95 (0.04) & 15.31 (0.02) & --- & 1.17 & 0.95 & 0.67 & 0.43\\
WS5$^*$      & $  2.89780   \pm    0.00003 $ & 16.71 (0.06) & 16.89 (0.03) &
16.32 (0.02) & 15.63 (0.03) & --- & 1.11 & 0.70 & 0.42 & 0.30\\
new      & $  2.94293   \pm    0.00002 $ & --- & 16.80 (0.02) &
16.17 (0.02) & 15.42 (0.03) & --- & --- & 0.72 & 0.47 & 0.27 \\
HV12203  & $  2.95411   \pm    0.00002 $ & 17.03 (0.02)& 16.77 (0.03) &
16.14 (0.02)& 15.42 (0.02) & 14.58 & 1.13 & 0.88 & 0.61 & 0.39 \\
We8      & $  3.039849  \pm   0.000001 $ & 17.04 (0.09)& 16.79 (0.04) &
16.14 (0.02) & 15.41 (0.03) & 14.52 & 0.92 & 0.74 & 0.50 & 0.33 \\
We3      & $  3.04904   \pm    0.00002 $ & 16.74 (0.06)& 16.54 (0.01) &
15.99 (0.01) & 15.32 (0.01) & --- & 1.06 & 0.81 & 0.59 & 0.36\\
WS11     & $  3.05330   \pm    0.00002 $ & 16.91 (0.13)& 16.54 (0.02) &
16.00 (0.01) & 15.34 (0.02) & --- & 1.06 & 0.62 & 0.42 & 0.29 \\
We2      & $  3.05485   \pm    0.00002 $ & 16.77 (0.20) &16.59 (0.01) &
16.01 (0.01) & 15.31 (0.01) & 14.41 & 1.03 & 0.87 & 0.59 & 0.39\\
WS9$^*$       & $  3.06945   \pm    0.00002 $ & 16.04 (0.03) & 15.93 (0.06) &
15.59 (0.04) & 15.12 (0.02) & --- & 0.51 & 0.31 & 0.27 & 0.18 \\
V1       & $  3.08455   \pm   0.00001  $ & --- & 16.77 (0.02) &
16.13 (0.01) & 15.38 (0.01) & --- & --- & 0.74 & 0.50 & 0.37\\
HV12202  & $  3.10120   \pm   0.00003  $ & 16.99 (0.12) & 16.74 (0.03)
& 16.10 (0.01)& 15.37 (0.01)& 14.40 & 1.06 & 0.73 & 0.52 & 0.34\\
HV12197  & $  3.14371   \pm   0.00003  $ & --- & 16.76 (0.02)&
16.11 (0.02) & 15.37 (0.02) & 14.47 & --- & 0.79 & 0.52 & 0.35\\
We5$^*$       & $  3.1745   \pm   0.0001  $ & 16.19 (0.09)& 16.22 (0.04)&
15.75 (0.04)& 15.17 (0.02) & --- & 0.88 & 0.55 & 0.42 & 0.27 \\
We7$^*$       & $  3.23227   \pm   0.00002  $ & 16.11 (0.10) & 15.95 (0.07) &
15.54 (0.07) & 15.03 (0.05) & --- & 0.91 &  0.79 & 0.65 & 0.41\\
We6      & $  3.28994   \pm   0.00002  $ & 16.56 (0.06)& 16.59 (0.02)&
15.99 (0.02)& 15.29 (0.02)& --- & 0.54 & 0.48 & 0.32 & 0.22\\
V4       & $  3.318     \pm     0.001  $ & 17.06 (0.06)& 16.78 (0.03) &
16.10 (0.01)& 15.34 (0.01) & 14.39 & 0.44 & 0.39 & 0.25 & 0.17 \\
HV12204$^{**}$  & $ 3.43882   \pm   0.00002  $ & --- & 16.23 (0.02) &
15.72 (0.02) & 15.08 (0.02)& 14.26 &  1.32 & 0.92 & 0.67 & 0.60 \\
V7       & $  3.45207   \pm   0.00001  $ & 16.91 (0.06) & 16.64 (0.02) &
16.00 (0.01) & 15.27 (0.02) & 14.32 & 0.69 & 0.47 & 0.32 & 0.22\\
HV12198  & $  3.522800  \pm  0.000006  $ & 16.88 (0.07) & 16.64 (0.03) &
15.98 (0.01) & 15.23 (0.03) & 14.32 & 1.26 & 0.92 & 0.63 & 0.36\\
\hline
\multicolumn{11}{l}{\footnotesize{$^*$  These Cepheids have scattered light curves and are not
considered in the following analysis.}}\\
\multicolumn{11}{l}{\footnotesize{$^{**}$  The membership to NGC1866
    of these  Cepheids was ruled out by \citet{Welch91} and they are not
considered in the following analysis.}}
\end{tabular}
\end{center}
\end{table*}

Analyzing the Cepheid light curves in Fig. \ref{fig:lcurves}, we note
that We5, We7 and WS9 appear to be affected by more noise than
the other objects, in particular in the {\it B\/} and {\it V\/} bands, and we can see 
similar but less evident scatter also for We4 and
WS5. This is likely the result of blending due to the location of these variables 
in the central very crowded region, combined with varying seeing conditions.
Moreover, a time-series
analysis for double-mode behaviour for such crowded stars is likely inconclusive given the long intervals between observational epochs and the relatively short periods with excellent seeing. 
The Cepheids
in the external region have well defined light curves.  V5,
V6 and V8 have periods and light curves typical of First
Overtone (FO) Cepheids. In particular, V5 is the brightest Cepheid and has
a very blue color but, as pointed out by \citet{Welch91}, its
membership seems to be excluded according to its mean radial velocity and  large
distance from the center of the cluster. \citet{Welch91}, on the basis
of the radial velocity, also ruled out membership for HV12204.


\begin{figure*}
\includegraphics[width=0.9\textwidth]{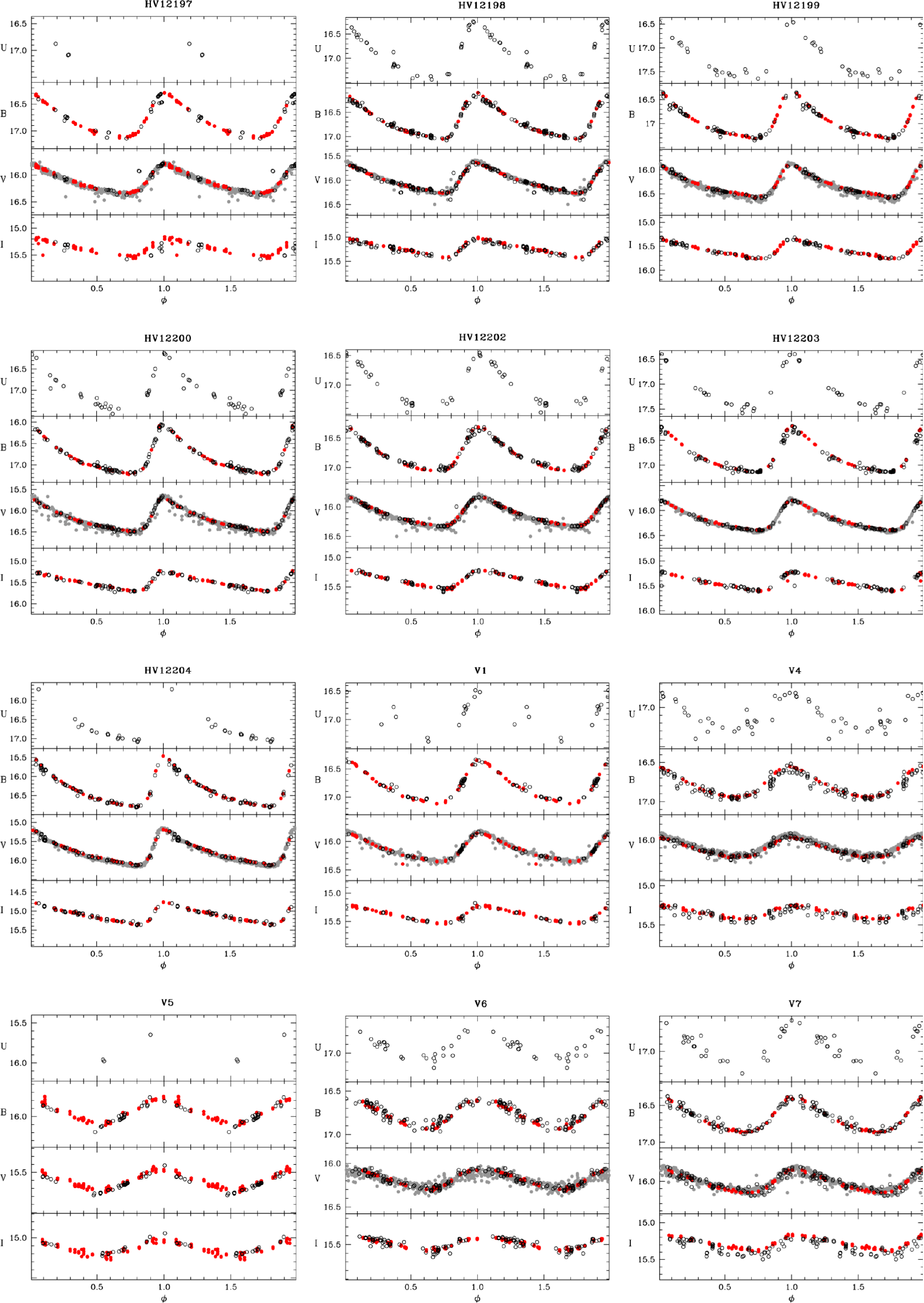} 
\caption{Light
curves of NGC1866 Cepheids. Grey dots are macho data, red dots are VLT
data and open circles
are the other archival data (see text for details). }
\label{fig:lcurves}
\end{figure*}

\begin{figure*}
\includegraphics[width=0.9\textwidth]{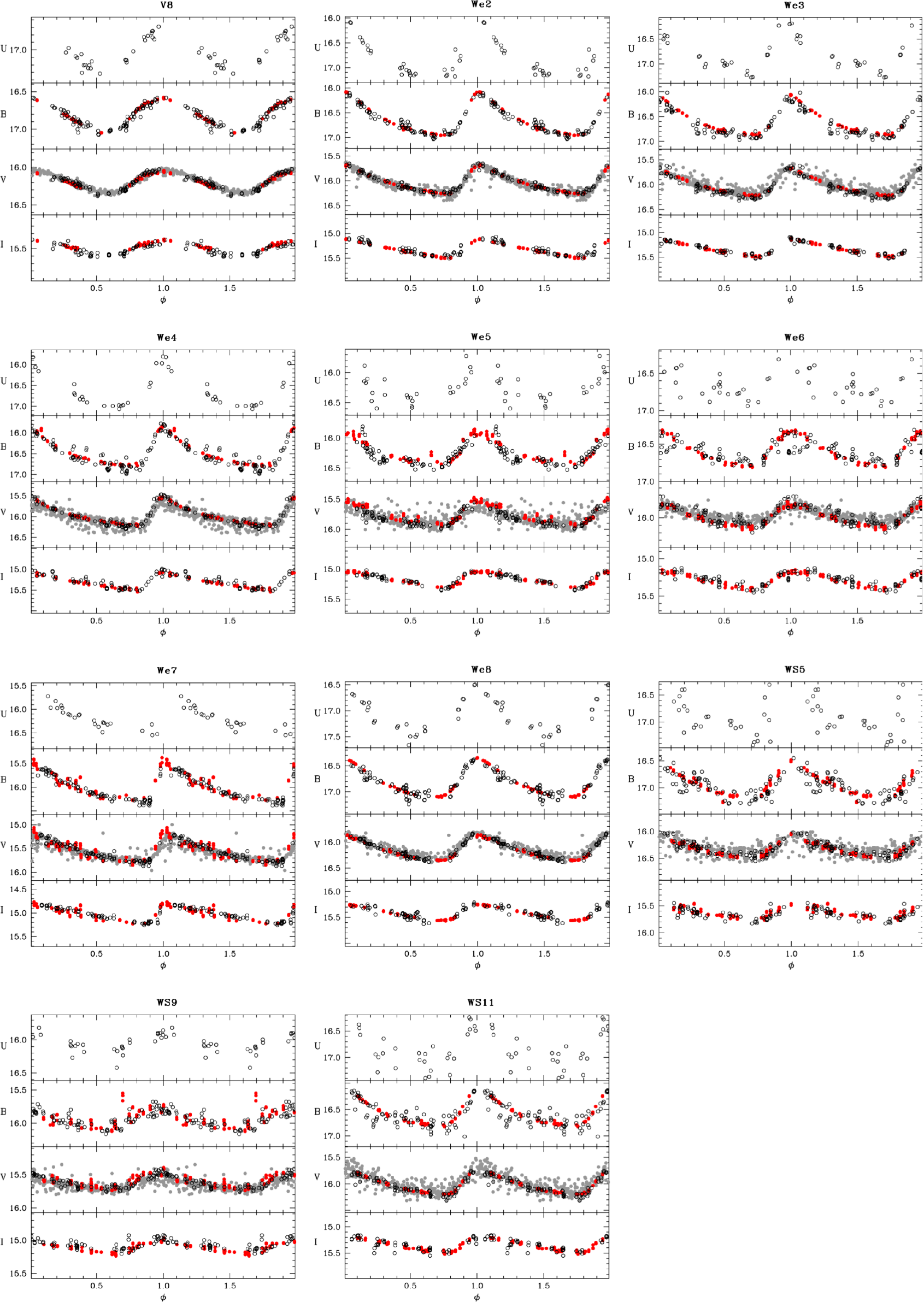} 
\contcaption{}
\end{figure*}



\section{Pulsation Models} \label{models}

For the interpretation of
the observed properties of the variable stars in NGC1866, we adopt 
the theoretical pulsation scenario developed by our team \citep[see
e.g.][and references therein]{bms99,Marconi05,Marconi10,Marconi13} and
based on nonlinear convective pulsation models for different
assumptions on the metallicity.
For each assumed fixed chemical composition, a wide range of model masses is
explored and the luminosities are selected according to evolutionary
predictions for the mass-luminosity (ML) relation of central helium-burning
intermediate-mass stars. In the ``canonical'' scenario
 both mass loss and convective overshooting are neglected, whereas in
 the ``noncanonical'' assumption an overluminosity of 0.25 dex at
 fixed mass is expected as a result of mild core overshooting and/or
 moderate mass loss \citep[see e.g.][for details]{Chiosi93,bccm99}. 
According to this theoretical scenario the effect of variations in both the
metallicity and the helium content on the predicted coefficients of the PL
and PLC relations is not negligible \citep{bccm99,Caputo00,Marconi05}.
On the other hand, we can adopt the Wesenheit (WPL) relations
\citep{Madore82}, defined as
$WPL(\lambda_1,\lambda_2)=\lambda_1-R_{\lambda_1\lambda_2} \times
(\lambda_1-\lambda_2)$, where $\lambda_i$ are the bands used and $R_{\lambda_1\lambda_2}=A_{\lambda_1}/E(\lambda_1-\lambda_2)$ is the ratio
of total to selective absorption \citep[see e.g.][]{Cardelli89}; this
represents a reddening-free formulation of the PL relation with a
reduced intrinsic scatter. In particular, \citet{Caputo00} showed that
the intrinsic scatter and the dependence on the
metallicity of the WPL({\it V,I\/}) relation is almost negligible.

 The coefficients of the theoretical PL and WPL
 relations for the chemical composition of NGC1866 ($Z=0.008$, $Y=0.25$), derived in the period range from $\log{P}=0.4$ (corresponding to the
 minimum fundamental period in NGC1866\footnote{We did not include
   shorter periods due to the expected deviation from linearity of the instability strip topology for models with
   mass lower than $\sim$ 4$M_{\odot}$ \citep[see][for details]{Bono01}}) to $\log{P}=1.5$ (corresponding to the maximum
 period in the OGLE sample), are those by \citet{Fiorentino02} and
 \citet{Fiorentino07}, respectively. In particular the WPL
 relation was also derived with an explicit dependence on both the
 metallicity and the
 excess luminosity $\log{L/L_c}$, where $L_c$ is the canonical
 luminosity level. This allows us to check the theoretical models 
both for \rm{canonical}  ($\log{L/L_c}=0$) and
 \rm{noncanonical} ($\log{L/L_c}=0.25$ dex)  assumptions \citep[see Table 3 in][]{Fiorentino07}. For the {\it V\/}, {\it I\/} filter combination, moving from
 the canonical to the non canonical assumption, the WPL({\it V,I\/})
 relation
 gets fainter by about 0.21 mag.

\section{The PL and Period-Wesenheit relations} \label{relations}

\begin{figure}
\includegraphics[width=8.5cm]{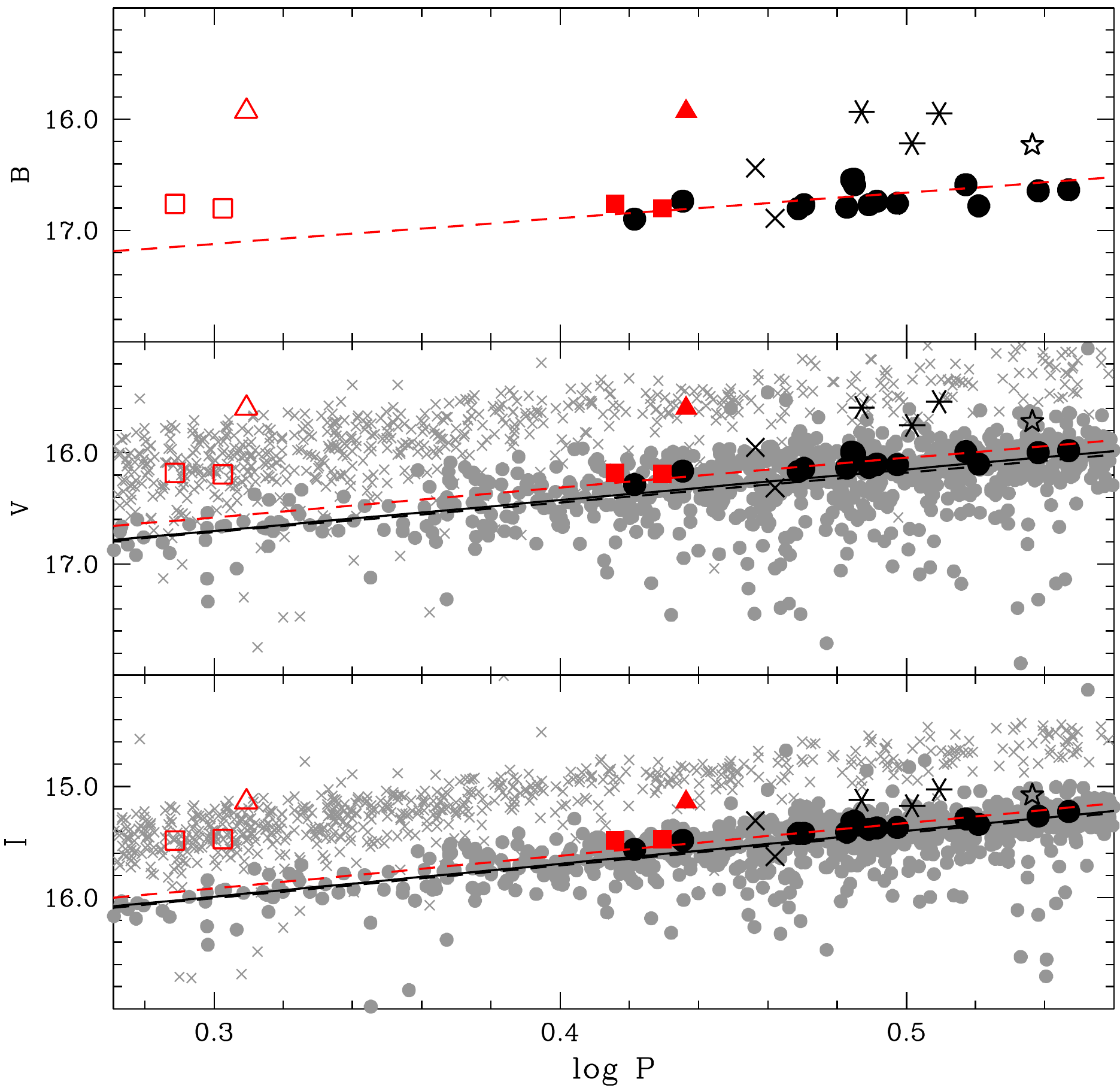} 
\caption{PL relation for NGC1866 Cepheids in the {\it B\/}, {\it V\/}
  and {\it I\/} bands. In
  the {\it V\/} and {\it I\/} bands, our sample is compared with the OGLE one. See
  text for symbols.} \label{PL}
\end{figure}

Fig. \ref{PL} shows the {\it BVI\/} PL relations for the Cepheids
 identified here
(black filled circles are the reliable fundamental Cepheids, asterisks
are We5, We7 and WS9,
crosses are We4 and WS5 and open star is HV12204). The red 
symbols represent the FO Cepheids (the triangle is V5 and
squares are V6
and V8, the two reliable FO Cepheids), plotted adopting both
their own (open red symbols) and their fundamentalised
periods\footnote{The period the variable would have if it were a
  fundamental one, obtained as $\log P_{fun}= \log P + 0.127$.} 
(filled red symbols).  It is worth noting that at fixed period, even taking into
account only the reliable Cepheids, the  scatter in magnitude is
significant (0.12, 0.09 and 0.07 mag in the $B$, $V$, and $I$ bands, respectively). Given that our targets share the same
age and chemical composition, 
this
 can only be ascribed to i)~the finite width of the instability strip \citep[see][for a
detailed discussion]{bccm99,Caputo00} and ii)~a possible mass-loss effect. 
For the {\it V\/} and {\it I\/} PL relations (middle and bottom panel of Fig. \ref{PL}), we
compared our sample
with the OGLE one (grey crosses are first overtone Cepheids and grey
filled circles  the fundamental ones). The black straight lines represent the
mean OGLE PL relations.  It is evident that the NGC1866 sample appears to be brighter than 
the OGLE III one \citep{Sosz08}.
Indeed, when we overlay the OGLE III {\it V\/} and {\it I\/} PL slopes on the Cepheid population
 of NGC1866\footnote{We notice that in all our fits of the NGC1866 sample, we include only the reliable fundamental Cepheids
and the two first overtone V6 and V8, adopting their fundamentalised
periods} we
obtain a shift of $\Delta\mu_V = -0.12 \pm 0.02$ mag and $\Delta \mu_I=-0.08\pm 0.02$ mag. 

The black dotted lines and the red dashed lines in Fig. \ref{PL} represent the
linear regression of the OGLE
data and of our Cepheid sample, respectively, adopting the theoretical
slopes for $Z=0.008$ in the \rm{canonical}  assumption and in the same
period range covered by OGLE  ($0.4<\log P<1.5$) reported in
 Table 6 of \citet{Fiorentino02}:
$$\overline{M_B} = -2.44 \log{P} -0.93$$ 
$$\overline{M_V} = -2.75 \log{P} -1.37$$ 
$$\overline{M_I} = -2.98 \log{P} -1.95$$


Adopting these theoretical slopes we
find for the OGLE sample $\mu_{V}^{OGLE}=18.904 \pm 0.007$ mag, $\mu_{I}^{OGLE}=
18.826 \pm 0.005$ mag  and for the NGC1866 sample (only reliable Cepheids)
$\mu_{B}^{NGC1866}=18.79\pm 0.03$ mag, $\mu_{V}^{NGC1866}=18.77 \pm
0.02$ mag, $\mu_{I}^{NGC1866}=18.74\pm 0.02$ mag (the adopted errors
are the standard deviation of the mean). 
To derive the absolute
distance moduli, we need to know the color excess. 
One of the  most used values for the color excess of NGC1866 and LMC is $E(B-V)=0.06$ mag \citep[see e.g.][]{Storm05,Schlegel98}. However, we
can also derive this value 
from the difference of the apparent moduli in two
different bands. In particular $E(B-V) =\mu_B - \mu_V$ and $E(V-I) =
\mu_V - \mu_I$. Adopting the inferred apparent distance moduli, we
obtain $E(V-I)_{OGLE}=0.078\pm0.009$ mag---a value that, combined with the extinction
law by \citet{Cardelli89}, gives $E(B-V)_{OGLE}=0.064\pm 0.007$
mag. From the apparent distance moduli of NGC1866, we obtain
$E(B-V)_{NGC1866}=0.02\pm0.04$ mag and
$E(V-I)_{NGC1866} = 0.03\pm0.03$ mag. 
The estimate for the LMC agrees quite well with the aforementioned
value $E(B-V)=0.06$ mag. However, the
values that we obtain for NGC1866 are significantly lower than the
expected ones, but the distribution of the Cepheid individual color
excesses is not Gaussian and if we compute the medians, we obtain
$E(B-V)_{NGC1866}=0.04\pm0.03$ mag and
$E(V-I)_{NGC1866} = 0.04\pm0.02$ mag. Moreover, it is worth
remembering that all the other NGC1866 stars that are bright enough to
be significant when contaminating the Cepheids' photometry (either as
binaries or due to coincidence along the line of sight) are going to
be bluer than the Cepheids, thus affecting
 {\it B\/} more than {\it V\/} and {\it V\/} more than {\it I\/}. Perhaps only the very
 outer stars could be contaminated by redder field stars. For this reason,
the reddening values determined by the apparent moduli differences are systematically underestimated.
Due to the strength of the literature color excess determination, we
adopt for both the samples 
the value $E(B-V) =0.06$ mag together with a ratio between $A_V$
and $E(B-V)$ given by $R_V=3.1$ \citep{Cardelli89}, obtaining
$\mu_{0}^{OGLE}=18.718$ mag and $\mu_{0}^{NGC1866}=18.58$ mag.

\begin{figure}
\includegraphics[width=8.5cm]{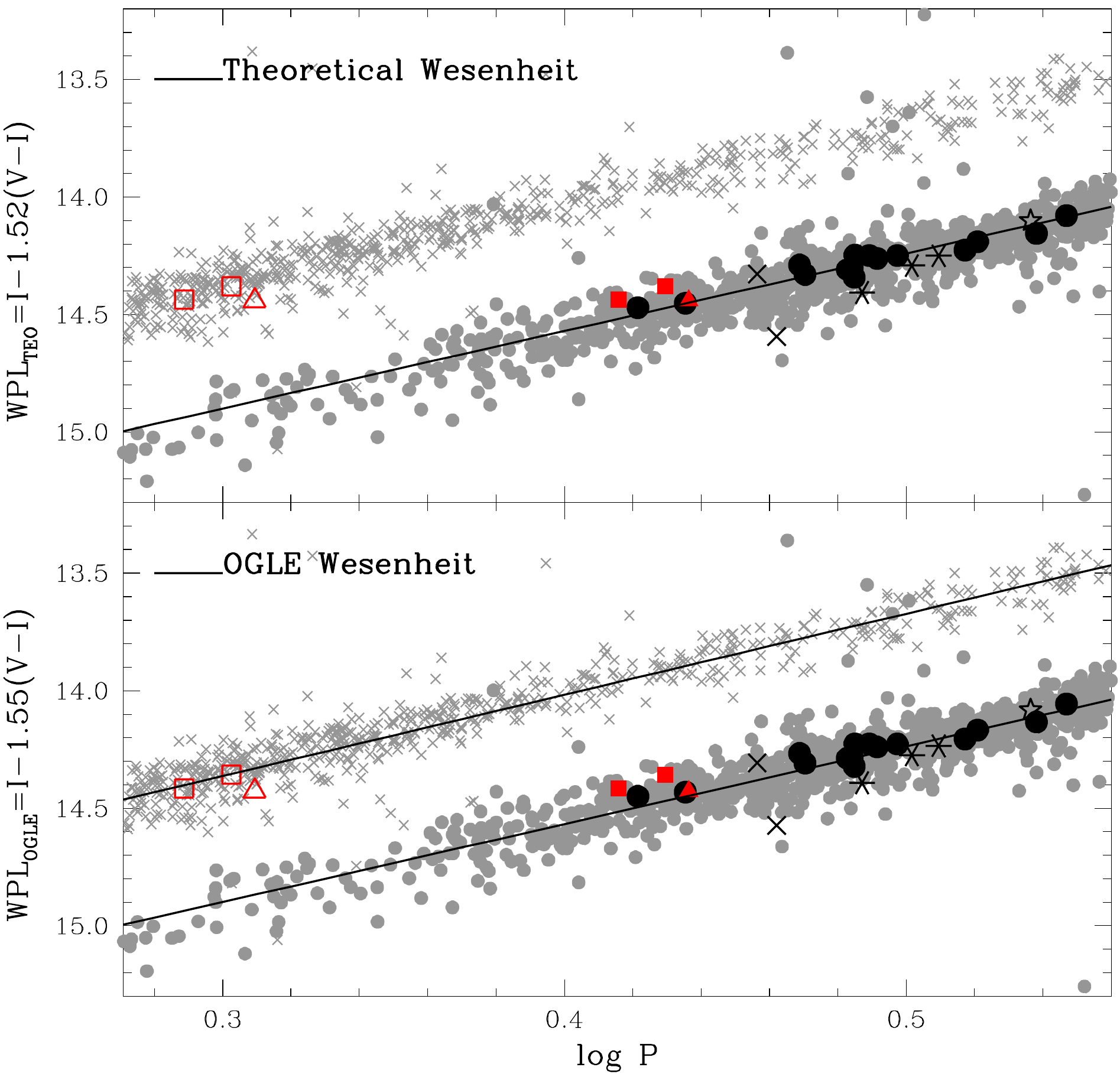} 
\caption{WPL($V,I$) relation of NGC1866 Cepheid sample
  compared with the OGLE one. In the upper panel we plot the WPL relation with
  1.52 as the color term to use the theoretical slope by
  \citet{Fiorentino07}. In the bottom panel, we adopt the OGLE WPL relation with
  1.55 as the color term.} \label{wes}
\end{figure}

To reduce the problems due to the determination of the color excess, we adopt the reddening free WPL
relations. It is worth noting that the definition of the WPL relation is not
unique: it depends on the adopted reddening law from which we derive
the color term.  
As theoretical  WPL relations, we adopt those, dependent on metallicity {\it Z\/} and luminosity level $\log
L/L_{\odot}$, obtained by \citet{Fiorentino07} (see their Table
3) using the absorption to reddening ratio from  \citet{Dean78} and
\citet{Laney93}. For the {\it V\/}and {\it I\/} bands, $WPL(V,I)=I-1.52(V-I)=
-2.67-3.30 \log P +0.84 \log L/L_{\odot}+0.08 \log Z$. On the other
hand, 
\citet{Sosz08} have derived an empirical OGLE-based relation adopting $WPL(V,I)=I-1.55(V-I)$. For this reason, we
cannot perform a direct comparison. In Fig. \ref{wes} in the upper
panel we plot WPL({\it V,I\/}) with 1.52 as color term and compare OGLE
and NGC1866 data with the theoretical relation in the \rm{canonical} assumption. In the bottom panel,
we adopt the WPL relation with 1.55 as color term and compare the data with the
empirical relation by OGLE. Symbols are the same as in
Fig. \ref{PL}. If we adopt the OGLE WPL({\it V,I\/}) relation, the difference
between the distance moduli of NGC1866 and LMC is $\Delta\mu =
-0.018 \pm 0.014$ mag. 
Moreover, using the slope of the plotted \rm{canonical}  WPL({\it V,I\/}) relation, we obtain $\mu_{0}^{OGLE}=18.717\pm 0.003$ mag and
$\mu_{0}^{NGC1866}=18.71 \pm 0.01$ mag. This result implies that removing
the reddening problem, the inferred LMC and NGC
1866 moduli are in good agreement. The values based on
the predicted \rm{canonical} WPL relation seem to favor  the ``long
distance scale'' for the LMC. On the other hand, as noted above, in
the \rm{noncanonical} assumption ($\log{L/L_c}=0.25$ dex),  the
predicted Wesenheit function is fainter by about 0.21 mag and in turn
the inferred  distance moduli  are $\mu_{0}^{OGLE}=18.507\pm 0.003$ mag and
$\mu_{0}^{NGC1866}=18.50 \pm 0.01$ mag. 


\section{The empirical route to the distance of NGC1866} \label{distance}

To test the results obtained with the theoretical relation, we discuss
two empirical calibrations of Cepheid distances based on two direct
methods: i)~the trigonometric parallaxes available for a sample of
Galactic Cepheids and obtained with the {\em Hipparcos} satellite
\citep{vanl07} and the HST \citep{Benedict07,Riess14,Casertano15}; ii)~the Baade-Wesselink
method applied to 70 Galactic \citep{Storm11mw} and 36 LMC Cepheids
\citep{Storm11lmc}.

 Following the  same procedure adopted in \citet{Ripepi12}, their sample
of MW Cepheids with distances from trigonometric parallaxes by
\citet{vanl07} and  \citet{Benedict07}
has been selected  (see their Table 6) retaining only the variables with $\delta \pi/\pi
\le 0.2$. For  variables present in both the Hipparcos and the HST
sets,  \citet{Ripepi12}  have used a weighted average of the two parallaxes. 
  Moreover, we add two Cepheids, SY Aur and SS CMa,  with very
  accurate parallaxes obtained 
  by \citet{Riess14} and \citet{Casertano15}, respectively. In Table
\ref{parall}, we report the parallax, the $V$ and $I$ magnitudes
\citep[taken from Table A1 in][]{vanl07} and the
Lutz-Kelker (LK) correction \citep[calculated according to][]{Benedict07} for the Galactic Cepheids used in our analysis.

\begin{table}
\caption{Galactic Cepheids with known parallaxes used to test the
  calibration of Cepheid distances obtained by mean of the theoretical
  WPL($V,I$) relation.} \label{parall}
\begin{center}
\begin{tabular}{l|c|c|c|c|c|c}
\hline
\hline
\noalign{\smallskip}
ID & $log P$  & $\pi$ & $\sigma_{\pi}$ & {\it V\/} & {\it I\/} & LK \\
 & [d]  & [mas] & [mas] & [mag] &[mag] & [mag]  \\
\noalign{\smallskip}
\hline
\noalign{\smallskip}

   $\beta$ Dor &      0.993      &   3.26 &   0.14 &   3.757 &  2.929  &$-$0.0184\\   
   RT Aur       &        0.572        & 2.40   & 0.19  &  5.448  & 4.811   &$-$0.0627\\
   $\zeta$ Gem  &    1.006        & 2.74  &  0.12   & 3.915  & 3.070   &$-$0.0192\\
   $\ell$ Car     &        1.551          & 2.03 &   0.16   &  3.698 &2.522   &$-$0.0621\\
   X Sgr        &           0.846         &  3.17  &  0.14    & 4.564 &3.635  &$-$0.0195\\
   W Sgr        &          0.880       &  2.30  &  0.19    & 4.670  & 3.834  &$-$0.0682\\ 
   FF Aql       &          0.650        & 2.64  &  0.16   &  5.373  & 4.513   &$-$0.0367\\
   $\delta$ Cep &     0.730        & 3.71  &  0.12    & 3.953  & 3.200  &$-$0.0105\\
   SS CMa &  1.092 &0.428 & 0.054 & 9.925 & 8.470 & $-$0.1192 \\
   SY Aur &   1.006 & 0.348 & 0.038  & 9.066 & 7.854 & $-$0.1592  \\
\hline
\end{tabular}
\end{center}
\end{table}

\begin{figure}
\includegraphics[width=8.5cm]{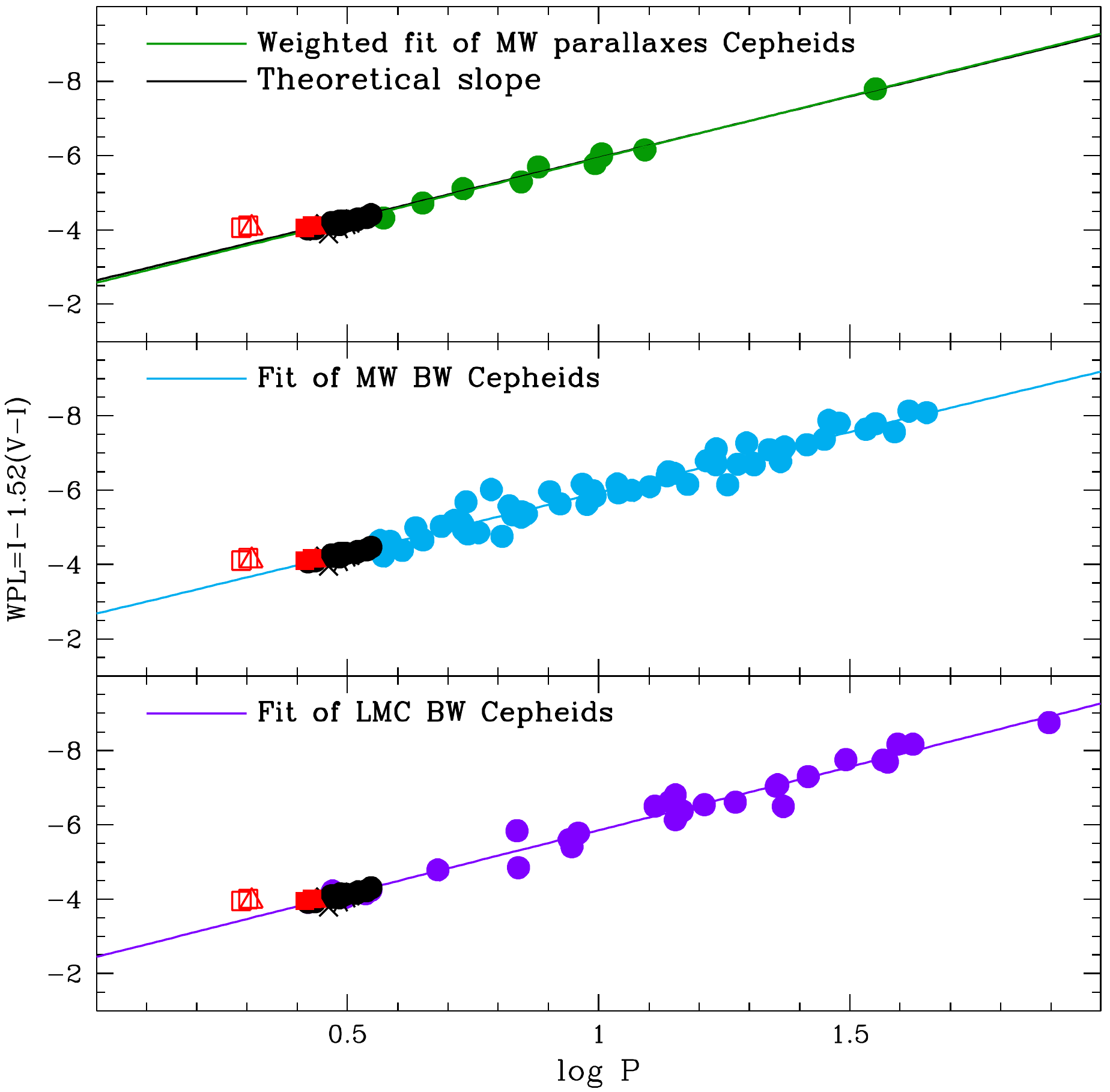} 
\caption{WPL({\it V,I\/}) relation for NGC1866 Cepheids (symbols used are
  the same of Fig. \ref{PL}) compared with different samples in
  literature. In particular, in the
  upper panel we adopt  Galactic Cepheids (green
  symbols) with
  distances obtained from parallaxes \citep{vanl07,Benedict07} and in the
  middle and bottom panels, we show
  Cepheids with  distances obtained from BW in the Galaxy \citep[cyan symbols][]{Storm11mw} and in
  LMC \citep[violet symbols][]{Storm11lmc}, respectively (see text for details). } \label{wconf}
\end{figure}

In the upper panel of Fig. \ref{wconf},
we show WPL({\it V,I\/}) (with the theoretical color coefficient of 1.52,
see above) for these Galactic Cepheids (green symbols) and for our sample (black and red symbols). The green line
represents the relation derived from a
weighted fit of the Galactic Cepheids absolute magnitudes, $WPL _{par} (V,I) =
-3.35 \log P -2.57$ ($\sigma=0.05$), whereas the black line is the relation obtained
assuming the slope of
the theoretical WPL({\it V,I\/}) relation for $Z=0.008$ discussed in the
previous section.  From the former relation we obtain
$\mu_{0}^{NGC1866} = 18.49 \pm 0.05$ mag, whereas  the theoretical relation
provides $\mu_{0}^{NGC1866} = 18.73 \pm 0.04$ mag in the \rm{canonical}
assumption and  $18.52 \pm 0.04$ mag in the \rm{noncanonical} one.  We note that
in spite of the difference between the slope obtained for the Galactic
Cepheids ($-3.35$)  and the theoretical one ($-3.30$),  when the theoretical
slope is applied to the Galactic Cepheids, the zero point
 obtained 
($-2.62 \pm 0.04$) is  in  good agreement with the
theoretical \rm{noncanonical} one ($-2.63$). 

 The middle and bottom panels of Fig. \ref{wconf}  show our
  NGC1866  data
(black and red symbols) together with MW
(cyan symbols) and LMC  (violet symbols) Cepheid samples with Baade-Wesselink (BW) distance estimates
by \citet{Storm11mw} and \citet{Storm11lmc}, respectively. In these
papers, the authors adopt 1.54 as color term in the WPL({\it V,I\/}).
For the Galactic sample, \citet{Storm11mw} give a WPL slope of
$-3.26$ and a zero point of $-2.7$. Adopting 1.52 as color term, we obtain
the same slope and a slightly different zero point of $-2.68$ (instead of the values $-3.38$  and
$-2.54$ obtained above from the weighted fit of the MW Cepheid with
parallaxes). The BW slope ($-3.26$) and the theoretical one ($-3.30$) are
very similar and when applied to the same NGC1866 sample, adopting
the zero point by \citet[][recomputed using 1.52 as color term coefficient]{Storm11mw}, provide distance moduli of 
$18.55 \pm 0.02$ mag and $18.53 \pm 0.04$ mag, respectively.

\begin{figure}
\includegraphics[width=8.5cm]{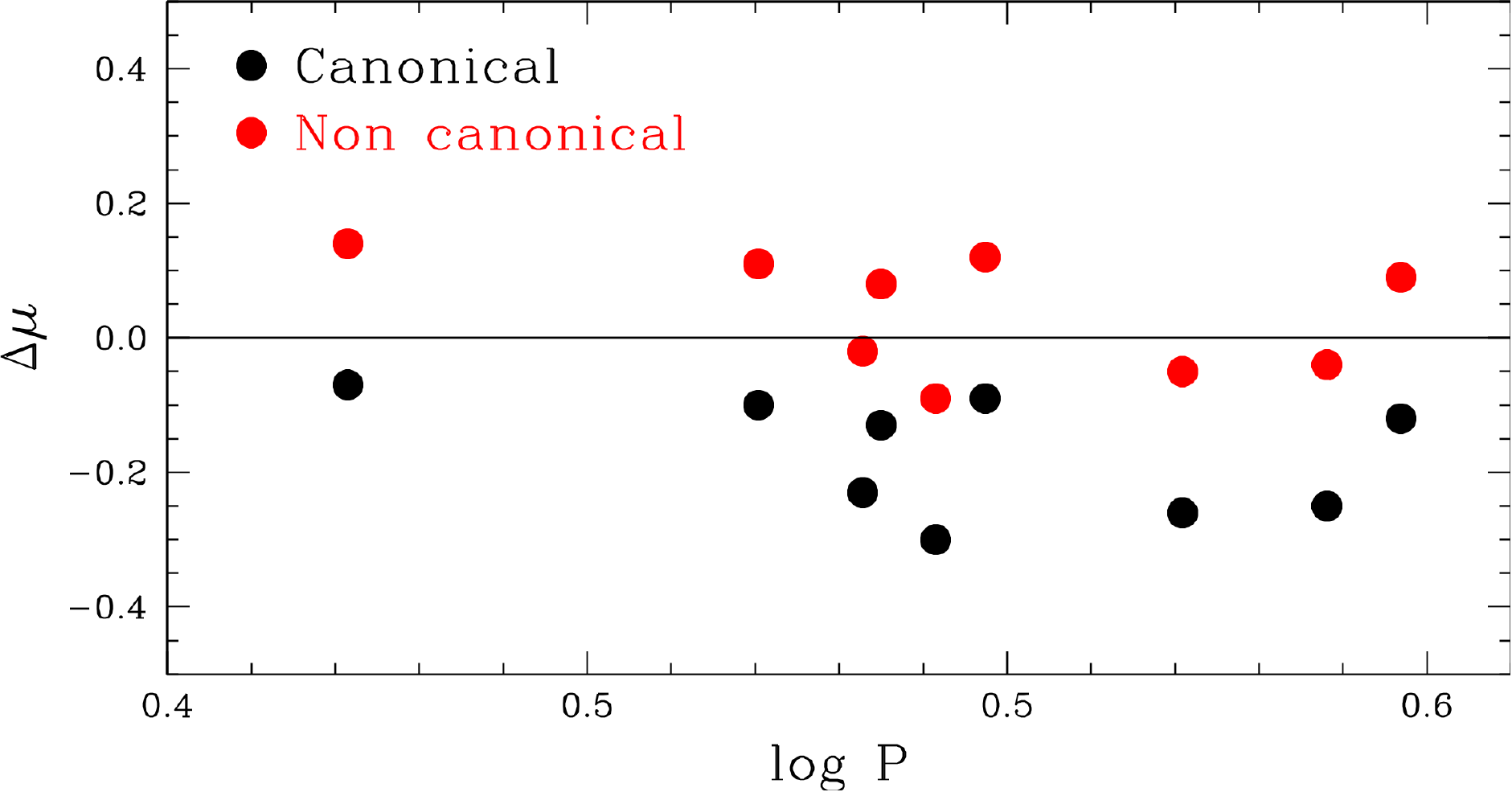} 
\caption{The differences between the
individual moduli obtained by \citet{Molinaro12} for a sample of 9
fundamental NGC1866 Cepheids (see their Table 5) and those obtained for the same
variables, adopting the theoretical slope and zero point of the WPL
for $Z=0.008$, both in the canonical (black symbols) and in the
non-canonical (red symbols) assumption. The black line corresponds to
$\Delta \mu =0$.} \label{confmolinaro}
\end{figure}

For the LMC BW Cepheids, \citet{Storm11mw}   give a WPL slope of $-3.41$
and a zero point of $-2.46$. Adopting a color term of 1.52, we obtain
the same slope and $-2.44$ as zero point. Applying this relation, we obtain a distance modulus 
of $18.39 \pm 0.03$ mag for NGC1866. 
The shorter distance scale obtained from the BW calibration supports
previous determinations based on this method in the literature \citep[see, e.g., the discussion
in][and references therein]{Molinaro12}. On the other hand the application of  the
CORS version of the BW method by \citet{Molinaro12} to the NGC1866 Cepheids used in their work gives
a distance modulus of $18.51 \pm 0.03$ mag. The uncertainties on the
distance estimates based on BW techniques depend on the controversial issue
of the adopted projection factor  which converts the measured radial
velocity into the pulsational one \citep[see e.g.][and references therein]{Barnes09,Storm11lmc,Molinaro12}. 
In Fig. \ref{confmolinaro}, we plot the differences between the
individual moduli obtained by \citet{Molinaro12} for a sample of 9
fundamental NGC1866 Cepheids and those obtained for the same
variables adopting the theoretical slope and zero point of the WPL
for $Z=0.008$, both in the canonical (black symbols) and in the
non-canonical (red symbols) assumption. In this figure we report also
a black line for $\Delta \mu =0$ mag corresponding to perfect
agreement. 

Inspection of these values and of Fig. \ref{confmolinaro} indicates that the adopted theoretical
scenario in the noncanonical assumption provides better agreement
with the most recent and widely assumed distance estimates for
NGC1866. Indeed, for the LMC, being near to our Galaxy and hosting several
stellar distance indicators, we have hundreds of different and
independent distance measurements. For example, a recent interesting 
measurement was obtained by the Carnegie Hubble Program \citep{Monson12}
to calibrate the extragalactic distance scale, with the final
  goal  to obtain $H_0$ with an
accuracy of 2\%. For this purpose, they 
observed Cepheids in
the Milky Way and nearby galaxies in the mid-infrared bands and found
for the LMC a modulus of $18.48\pm 0.04$ mag. The same value (with an
error of $\pm 0.05$ mag) was
obtained by \citet{Walker12} by the analysis of LMC distances
based on different stellar distance indicators (Cepheids, RR Lyrae,
Red Variables, Red Clump Stars and Eclipsing Variables). Finally it is
worth noting that \citet{Pietrzynski13} give a very accurate (within
2\%) LMC distance measurement based on the observations of 8 late-type
eclipsing binaries in this galaxy, finding a value of $18.493\pm
0.008$(statistical)$\pm0.047$(systematic) mag.

This conclusion seems to favor the hypothesis of a moderate
overluminosity of NGC1866 Cepheids at fixed mass, possibly due to a
combination of core overshooting and mass loss effects, in agreement
with the results by \citet{Brocato04}.

\section{The color-magnitude diagram} \label{cmd}

\begin{figure}
\includegraphics[width=8.5cm]{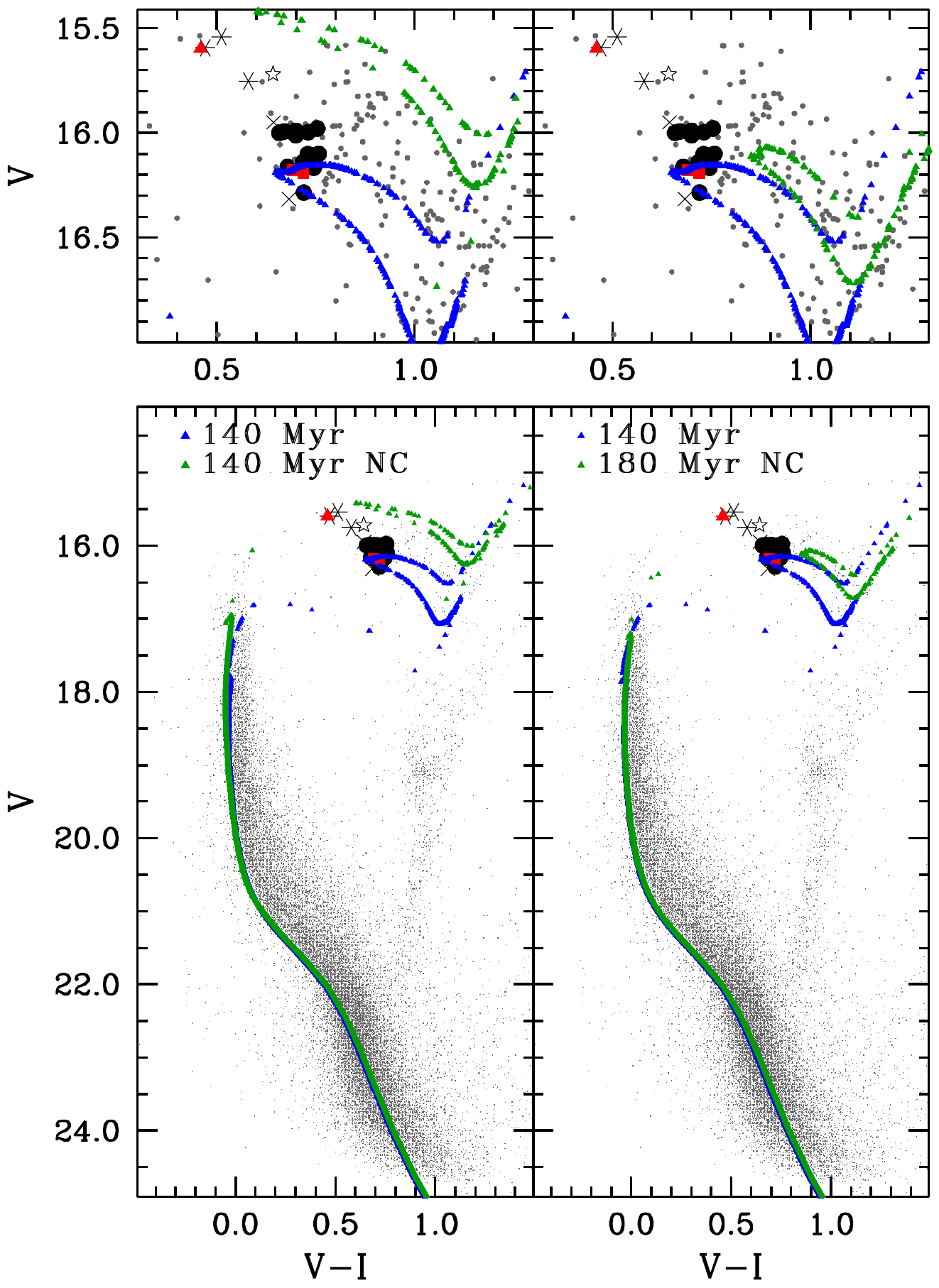}
\caption{In this figure we show, in the bottom panels, the {\it V,V-I\/} CMD of NGC1866 obtained using only the very
  precise VLT data compared with the synthetic CMDs computed with the
  stellar population synthesis code SPoT by \citet{Brocato03}. The
  symbols used for NGC1866 Cepheids are the same of Fig. \ref{PL}. The comparison with canonical (blue dots) and
noncanonical (green dots and labeled as NC) synthetic CMDs are shown. The age adopted
for the canonical synthetic CMD is 140 Myr, and that adopted for the
noncanonical one is 140 Myr in the left panel and 180 Myr in the right
one. In the top panels, there is a zoom of the region around the
Cepheid IS.} \label{cmd_sint}
\end{figure}

\begin{figure}
\includegraphics[width=8.5cm]{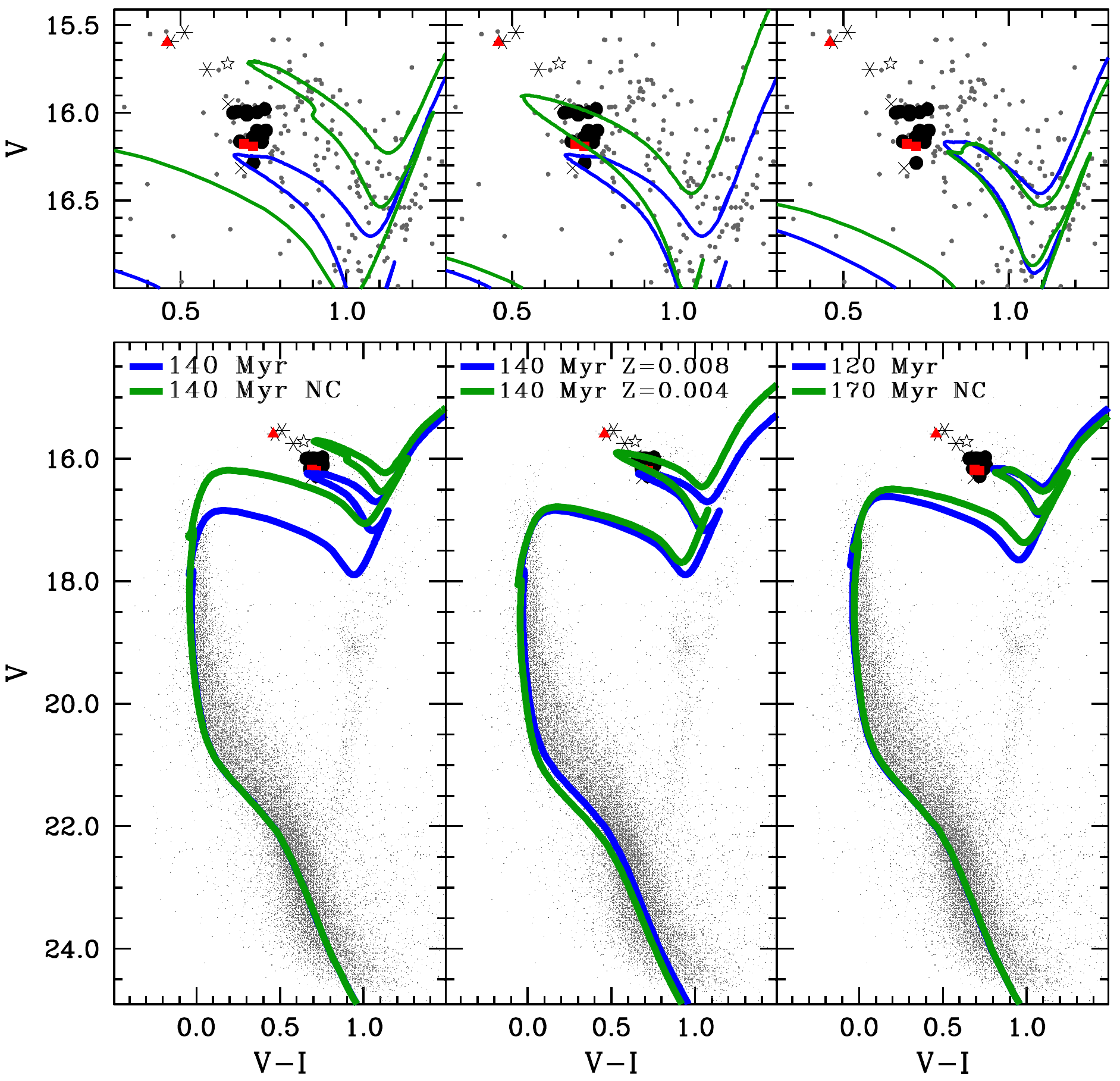}
\caption{In the three bottom panels of this figure, we show the same observational CMD of Fig. \ref{cmd_sint}
  compared with the isochrones from the BaSTI (Bag of Stellar Tracks
and Isochrones) database \citep{Pietrinferni04,Pietrinferni06} for
different assumptions of overshooting efficiency, metallicities and
ages (see text for details). In the top panels, there is a zoom of the region around the
Cepheid IS.} \label{cmd_isoc}
\end{figure}

As a by-product of the accurate time-series multiband observations, we
are able to build new 
CMDs. Figs. \ref{cmd_sint} and \ref{cmd_isoc} show in the bottom
panels the {\it V,V-I\/} CMD of NGC
1866 and in the top panels a zoom of the region around the Cepheid IS. To derive the mean magnitudes for the stars identified in the
observed field, we use only the VLT data that, as described in Sect. \ref{data}, are much more precise
than the other ones. Thanks to the robustness of estimated mean
magnitudes and colors, most of the reliable fundamental Cepheids
(black filled circles) are located in a
very restricted region of
the diagram, corresponding to the tip of the blue loop of He burning
giants. 

To compare the observed CMD and the Cepheid location with evolutionary
prescriptions, in Fig. \ref{cmd_sint} we considered the synthetic CMDs
computed with the stellar population synthesis code SPoT\footnote{For
  the details of the numerical procedures and physical assumptions of
  the SPoT code see \citet{Brocato00}, \citet{Raimondo05} and \citet{Raimondo09}.} by
\citet{Brocato03}, based on two sets of stellar evolutionary tracks calculated for this purpose
with the Pisa evolutionary code
\citep{Castellani03}, with canonical (no overshooting) and
non-canonical (moderate overshooting, 0.25 dex) assumptions. In particular, we considered  $Z =0.007$ and ages
of 140 Myr and 180 Myr for comparison with the results by
\citet{Brocato03,Brocato04} (see those papers for more
details). Moreover, we assumed a distance modulus of 18.5 mag and a color excess $E$({\it B-V\/}) = 0.06 mag. The
left panel of Fig. \ref{cmd_sint} shows the comparison between canonical and noncanonical
synthetic CMDs for an age of 140 Myr with the observed data.
We note that the location of the Cepheids is intermediate between
the blue loops obtained in the two assumptions. This suggests a possibly lower overshooting efficiency compared to that adopted, or a slightly lower cluster metallicity.
In the right panel, we use the same canonical synthetic CMD, but
increase the age of the noncanonical one to 180 Myr. We note that, for
the higher age, the luminosity level predicted in the noncanonical
assumption seems to reproduce the observed one, but the blue loop
extension is too short (see also discussion below). 

We also adopted the isochrones from the BaSTI (Bag of Stellar Tracks
and Isochrones) database \citep{Pietrinferni04,Pietrinferni06}, which
also allows  us to compare models based on the canonical assumptions and models including core overshooting at different chemical compositions. In the left panel of Fig.  \ref{cmd_isoc},  at fixed metallicity $Z =0.008$ and age of 140 Myr, we compare the canonical isochrone with the non-canonical one. As expected, the canonical blue loop is fainter than the observed Cepheids,
whereas the non-canonical one is too bright. We note that the
overshooting efficiency adopted in the BASTI database is almost twice
that adopted in the pulsational scenario, so that the
result obtained seems to confirm our finding based on the pulsational analysis
that Cepheids in NGC1866 might be affected by mild overshooting
of 0.2 dex and/or a small amount of mass loss.
At fixed age, the loop luminosity is also affected by metallicity, thus
in the central panel we compare the canonical isochrone at 140 Myr for $Z=0.008$
with the one for $Z=0.004$. The latter isochrone seems to better
match the observed Cepheid location, but it does not properly
reproduce the Main Sequence (MS). However, we cannot exclude an
intermediate metallicity close to $Z=0.006$.
Finally, in the right panel, we try to reproduce the luminosity of the
observed Cepheids by varying the isochrone age in the two
scenarios, assuming a metallicity of $Z=0.008$. We find that  in the canonical assumption the predicted
age is close to 120  Myr and in the non-canonical one it is about 170
Myr. The overshooting efficiency in the non-canonical isochrone is higher than expected according to the
pulsational results. In the case of mild overshooting,
an intermediate age of  about 140 Myr is predicted.
We note that the theoretical blue loops for $Z=0.008$ appear too
short at least for some ages and this discrepancy can be due to
different numerical and physical assumptions in the theoretical
scenario \citep{Bono00b,Castellani90,Valle09,Walm15}, even if---as
noted above---a slightly lower metal content cannot be excluded.

\section{Conclusions.} \label{conclusions}

We have investigated the properties of a sample of Cepheids in NGC1866 
on the basis of both archival and very precise VLT data.

As a result of the time series analysis,  accurate light curves
are obtained for 21 fundamental and three first overtone
pulsators. We also derive the mean magnitudes and colors for each
pulsator and compare their behavior with both empirical (OGLE) and predicted 
LMC PL and WPL({\it V,I\/}) relations.
The results from the PL are affected both by the intrinsic
dispersion and by reddening uncertainties, whereas the WPL({\it V,I\/}) relation
gives more reliable results. Adopting the slope of the theoretical \rm{canonical}  WPL({\it V,I\/}) relation, we obtain $\mu_{0}^{OGLE}=18.717\pm 0.003$ mag and
$\mu_{0}^{NGC1866}=18.71 \pm 0.01$ mag, for the OGLE LMC field sample and
our NGC1866 data, respectively. On the other hand in
the \rm{noncanonical} assumption ($\log{L/L_c}=0.25$ dex), 
the inferred  distance moduli are $\mu_{0}^{OGLE}=18.57\pm 0.003$ and
$\mu_{0}^{NGC1866}=18.50 \pm 0.01$ mag.

As a test of our theoretical approach, we also considered two empirical calibrations of Cepheid
distances based on trigonometric parallaxes and BW measurements. Even
if the slope of the WPL({\it V,I\/}) relation obtained for the
Galactic Cepheids with
parallaxes is steeper by 0.06 than the theoretical one, its
application to the NGC1866 sample gives a distance modulus of 
$18.49 \pm 0.07$ mag, in very good
agreement with that obtained applying the theoretical noncanonical slope ($18.52 \pm 0.04$ mag).

On the other hand, using Galactic Cepheids with BW measurements, the
slope of the WPL({\it V,I\/}) relation is very similar to the theoretical one. Both the
slopes give the same distance modulus of
18.54 mag for NGC1866, adopting the
corresponding zero point. 
Using the LMC BW sample and adopting
the \citet[][]{Storm11lmc}  WPL({\it V,I\/} relation (recomputed using 1.52 as color term), we obtain for NGC1866 a distance
of $18.39\pm 0.03$ mag---significantly shorter than the values derived above. This
shorter distance scale is in
agreement with previous determinations based on the BW method  and its
deviation from most of the other recent results for the LMC  has
already been debated in the literature \citep[see
e.g.][]{Molinaro12}. 

Therefore, our investigation seems to favor a value close to 18.5
mag for the distance modulus of NGC1866, as also obtained by the
theoretical noncanonical WPL({\it V,I\/}) relation, thus suggesting that mild
overshooting and/or moderate mass loss can affect intermediate-mass stellar evolution in
this cluster.

To conclude, we have compared the {\it V,V-I\/} CMD of NGC1866 with the predictions
of evolutionary and synthetic models, adopting $\mu_0=18.5$ mag,
$E(B-V)=0.06$ mag. The main conclusions of this comparison are:

\begin{itemize}
\item for the typical metallicity adopted for NGC1866 ($Z=0.007 \div
  0.008$), an age close to 120 Myr (170 Myr) is suggested  by the comparison
  with canonical (noncanonical)  isochrones.
 As the
  noncanonical isochrones are computed for an overshooting efficiency almost
twice that assumed in the pulsational analysis, we expect an age
around 140 Myr in the mild overshooting 
and/or moderate mass loss scenario;
\item decreasing the metallicity of the canonical isochrone at 140 Myr
  from $Z=0.008$ to $Z=0.004$, we find better agreement with the Cepheid
  distribution, but a worse fit for the main sequence phase. On this
  basis, an intermediate metallicity close to $Z=0.006$ cannot be
  ruled out.
\end{itemize}

Finally, at $Z=0.008$, taking into account the moderate overluminosity
required by pulsational models to find a distance modulus
in agreement with most empirical calibrations, we can conclude that our age determination for NGC1866 is close to 140 Myr.

\section*{Acknowledgments}

This paper is based on observations made with the ESO/VLT.
This work has made use of BaSTI web tools and was supported by PRIN-INAF 2014 “EXCALIBURS: EXtragalactic distance scale CALIBration Using first-
Rank Standard candles.” (PI: G. Clementini). 
 We thank our anonymous referee for her/his valuable comments.

\newpage

\bsp

\label{lastpage}

\end{document}